\documentclass{acm_proc_article-sp}
\pdfoutput=1
\usepackage{graphicx}
\graphicspath{{./images}}
\newcommand{\mcal}[1]{\mathcal{#1}}
\newcommand{\simil}{\mathop{\mathrm{sim}}}
\newcommand{\md}{{\rm d}}
\newtheorem{definition}{Definition}

\begin{document}

\title{Document Relevance Evaluation via Term Distribution Analysis Using Fourier Series Expansion}

\numberofauthors{3}
\author{
\alignauthor
Patricio Galeas\\
       \affaddr{Dept. of Math. \& Comp. Sci.}\\
	   \affaddr{University of Marburg}\\
	   \affaddr{Hans-Meerwein-Str. 3}\\
	   \affaddr{D-35032 Marburg, Germany}\\
       \email{galeas@informatik.uni-marburg.de}
\alignauthor 
Ralph Kretschmer\\
       \affaddr{Kretschmer Software}\\
       \affaddr{Zum Bernstein 3}\\
       \affaddr{D-57076 Siegen, Germany}\\
       \email{kretschmer@kretschmer-software.com}
\alignauthor
Bernd Freisleben\\
       \affaddr{Dept. of Math. \& Comp. Sci.}\\
	   \affaddr{University of Marburg}\\
	   \affaddr{Hans-Meerwein-Str. 3}\\
	   \affaddr{D-35032 Marburg, Germany}\\
       \email{freisleb@informatik.uni-marburg.de}
}

\maketitle

\begin{abstract}
In addition to the frequency of terms in a document collection, the distribution of terms plays an important role in determining the relevance of documents for a given search query. In this paper, term distribution analysis using Fourier series expansion as a novel approach for calculating an abstract representation of term positions in a document corpus is introduced. Based on this approach, two methods for improving the evaluation of document relevance are proposed: (a) a function-based ranking optimization representing a user defined document region, and (b) a query expansion technique based on overlapping the term distributions in the top-ranked documents. Experimental results demonstrate the effectiveness of the proposed approach in providing new possibilities for optimizing the retrieval process.
\end{abstract}


\section{Introduction}
One way to address the problems of synonymy and poly\-semy associated with lexical matching methods \cite{Berry94} in text retrieval applications is to consider contextual information \cite{Huang05}.
In fact, several search engines make use of contextual information to disambiguate query terms \cite{Lawrence00}. Contextual information is either obtained from the user, from the document structure or from the text itself by performing some form of statistical analysis, such as counting the frequency and/or distance of terms.

In this paper, a text retrieval approach that incorporates novel contextual analysis and document ranking methods is presented.
The proposed approach, called \textit{Fourier Vector Scoring}, is based on an abstract description of the term positions in a document, represented by the Fourier series expansion of a rectangular function describing the term positions in the document.
In addition, a document ranking optimization procedure, based on \textit{objective query functions} determining a user defined document region, is proposed as an alternative to the well-known term frequency  metrics.
Furthermore, a query expansion algorithm is introduced. It is based on overlapping the distributions of query terms in the top-ranked documents.
Experimental results obtained for the TREC-8 document collection demonstrate that the proposed approach is superior to state-of-the-art relevance feedback techniques such as \textit{Rocchio} and \textit{Divergence from Randomness} models \cite{Rocchio71,Amati02}.

The paper is organized as follows: Section 2 outlines related work on contextual information retrieval. In Section 3, term distribution analysis using Fourier series expansion is presented. The comparison of term distributions is described in Section 4. Section 5 discusses experimental results. Section 6 concludes the paper and outlines areas for future research.

\section{Related Work}
Contextual information can be obtained in two ways: by the text surrounding the search terms in the document corpus, or by the context delivered by the user (i.e.\ \textit{personalization}) \cite{Huang05}. There are approaches that utilize the query history of users \cite{Shen03} or the text surrounding the query \cite{Finkelstein02,Abdel-Razek03} to build augmented queries (i.e.\ \textit{query expansion}) for improving the performance of interactive retrieval systems.

Relevance feedback is the most popular \textit{query expansion} strategy \cite{SaltonBuckley90,Efthimiadis92,Buckley94}. Here, the expanded terms are typically extracted from the retrieved documents and judged as relevant in a previous retrieval iteration.
As demonstrated in several experimental studies, relevance feedback systems are quite effective \cite{Robertson76,Buckley95}. However, the browsing process required to determine the relevance of a document has been widely recognized as a significant limitation by the information retrieval research community.

To overcome the intervention of the user in the relevance feedback process, two basic types of strategies have been proposed: \textit{automatic global analysis} and \textit{automatic local analysis}.
In automatic global analysis, all documents of the collection are used to determine a \textit{thesaurus}-like structure, defining term-to-term relationships within the document corpus. In general, global analysis techniques are limited to small database applications, where doubtful improvements have been observed \cite{Attar81}.
In automatic local analysis, the system is able to estimate the relevance of the first retrieved documents without user intervention. The main idea is to consider the \textit{top-n} initially retrieved documents as relevant, and to use statistical heuristics to identify query related terms \cite{Efthimis93,Xu00}. \textit{Noise} and \textit{multiple topics} are two major negative factors for expansion term selection \cite{Yu03}. 
To deal with these problems, traditional clustering methods have been proposed \cite{Hearst96}. The experiments performed by Fan et al.\ \cite{Fan04} confirm that highly-tuned ranking offers more high-quality documents at the top of the hit list. 

Typically, it is difficult to determine correlated terms inside a document, 
because these terms do not necessarily co-occur very frequently with the original query terms if the document is considered as a whole.
In fact, it is common to have unrelated terms co-occurring with query terms very frequently \cite{Sun06}. To address this problem, page segmentation strategies have been suggested \cite{Yu03,Cai04}. They provide a better document partitioning at the semantic level and reduce the probability to carry irrelevant terms to the query expansion process.
In general, an important drawback of automatic local analysis strategies is the considerable amount of computation, which represents a substantial problem for interactive systems \cite{LI05}.

Katz \cite{Katz96} has analyzed the distribution of content-bearing terms in technical documents.  Important concepts supporting word occurrence models, such as \textit{inter-}/\textit{within-document} relationships, \textit{topicality} and \textit{burstiness} are proposed. The author concentrates on the modeling of the \textit{inter-document} distributions of content words, while our work focuses on the \textit{within-document} relationships applied to relevance evaluation in the information retrieval process.

An approach to apply term positional data in retrieval feedback is the work of Attar and Fraenkel \cite{Attar77}. They propose different models to generate clusters of terms related to a query (searchonyms) and use these clusters in a local feedback process. In their experiments with English and Hebrew documents, they confirm that metrical methods based on functions of the distance between terms are superior to methods based merely on weighted co-occurrences of terms.

Several approaches based on Hidden Markov Models have been proposed \cite{Miller99,Mittendorf94,Jinxi-Xu00}. For example, Miller et al.\ \cite{Miller99} propose a probabilistic model based on Bayes' theorem. Although this approach is mathematically elegant, its probabilistic hypothesis finally reduces it to a \textit{term frequency} representation.

One of the first approaches applying Fourier analysis to term distributions in documents is Fourier Domain Scoring (FDS), proposed by Park et al.\ \cite{Park04}. FDS performs a separate magnitude and phase analysis of term position signals to produce an optimized ranking. It creates an index based on page segmentation, storing term frequency and approximated positions in the document. FDS processes the indexed data using the \textit{Discrete Fourier Transform} to perform  the corresponding spectral analysis.
Our approach, on the other hand, represents the term signal information (Fourier coefficients) directly as an $n$-dimensional vector using the analytic Fourier transform,  thus permitting an immediate and simple term comparison process.

\section{Term Distribution Analysis Using Fourier Series}
Fourier analysis is based on the idea that functions can be approximated by a sum of sine and cosine waves at different frequencies. The more sinusoids are included in the sum, the better the approximation. 
There are several applications of Fourier analysis in the field of information retrieval (IR),
such as audio-IR \cite{Foote02}, image-IR \cite{Folkers02}, and in text-IR \cite{Park04}.

Consider a function $f(x)$ that is defined for $x \in [0, L]$. A Fourier series expansion is an expansion
\begin{equation}
f(x) = \frac{a_0}{\sqrt{L}} + \sqrt \frac{2}{L} \sum_{k = 1}^\infty
\left[a_k \cos\left(\frac{2 \pi k x}{L}\right) + b_k \sin\left(\frac{2 \pi k x}{L}\right)\right]
\label{fourier_expansion}
\end{equation}
where the coefficients $a_k$ and $b_k$ have to be determined.
If the sum over $k$ is restricted to $k \leq n$, the Fourier series gives an
approximation $f_n(x)$ to the function $f(x)$ called the $n$-th order
Fourier approximation of $f(x)$.

Consider a document $D$ containing $L$ terms. To characterize the distribution of a particular term $t$ within the document, the set of positions of all occurrences of $t$ in $D$ is denoted as $\mcal{P}_t$, where all terms are enumerated starting with $1$ for the first term in the document and so on.

\begin{figure}[ht]
	\centering
	\includegraphics[width=6cm,bb=0 0 225 162]{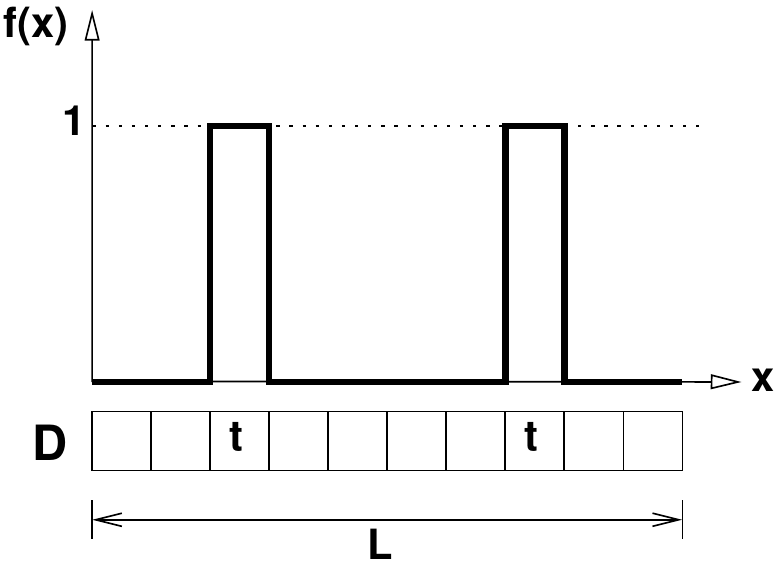}
	\caption{Distribution of the term \textit{t} in document $D$, represented by a rectangular function.}
	\label{pulse_distribution}
\end{figure}

As exemplified in Figure \ref{pulse_distribution}, $\mcal{P}_t = \{3, 8\}$ represents the fact that the two instances of the term $t$ in the document $D$ are located in the third and the eighth position of the document body.

The cardinality $|\mcal{P}_t|$ of $\mcal{P}_t$ is the total number of occurrences of $t$ in the document. The characteristic function
\begin{equation}
f^{(t)}(x) := \left\{ 
\begin{array}{ll}
1 & \mbox{for } x \in[p - 1, p] \mbox{ if } p \in \mcal{P}_t \\
0 & \mbox{otherwise}
\end{array} \right.
\end{equation} 
is assigned to $\mcal{P}_t$ for $x \in [0, L]$.
The Fourier coefficients of $f^{(t)}$ are given by
\begin{equation}
a_0 = \frac{|\mcal{P}_t|}{\sqrt{L}} 
\label{fourier_coef_0}
\end{equation}
and for $k >0$
\begin{equation}
a_k = \sqrt{ \frac{L}{2}} \frac{1}{k\pi} \sum_{p \in \mcal{P}_t} \left[
\sin \left( 2 \pi k \frac{p}{L} \right)
- \sin \left( 2 \pi k \frac{p-1}{L} \right) \right]
\end{equation}
\begin{equation}
b_k =  - \sqrt{\frac{L}{2}} \frac{1}{k\pi} \sum_{p \in \mcal{P}_t} \left[
\cos \left( 2 \pi k \frac{p}{L} \right)
- \cos \left( 2 \pi k \frac{p-1}{L} \right) \right]
\end{equation}

\begin{figure}[ht]
	\centering
	\includegraphics[width=1.0\linewidth,bb=0 0 350 242]{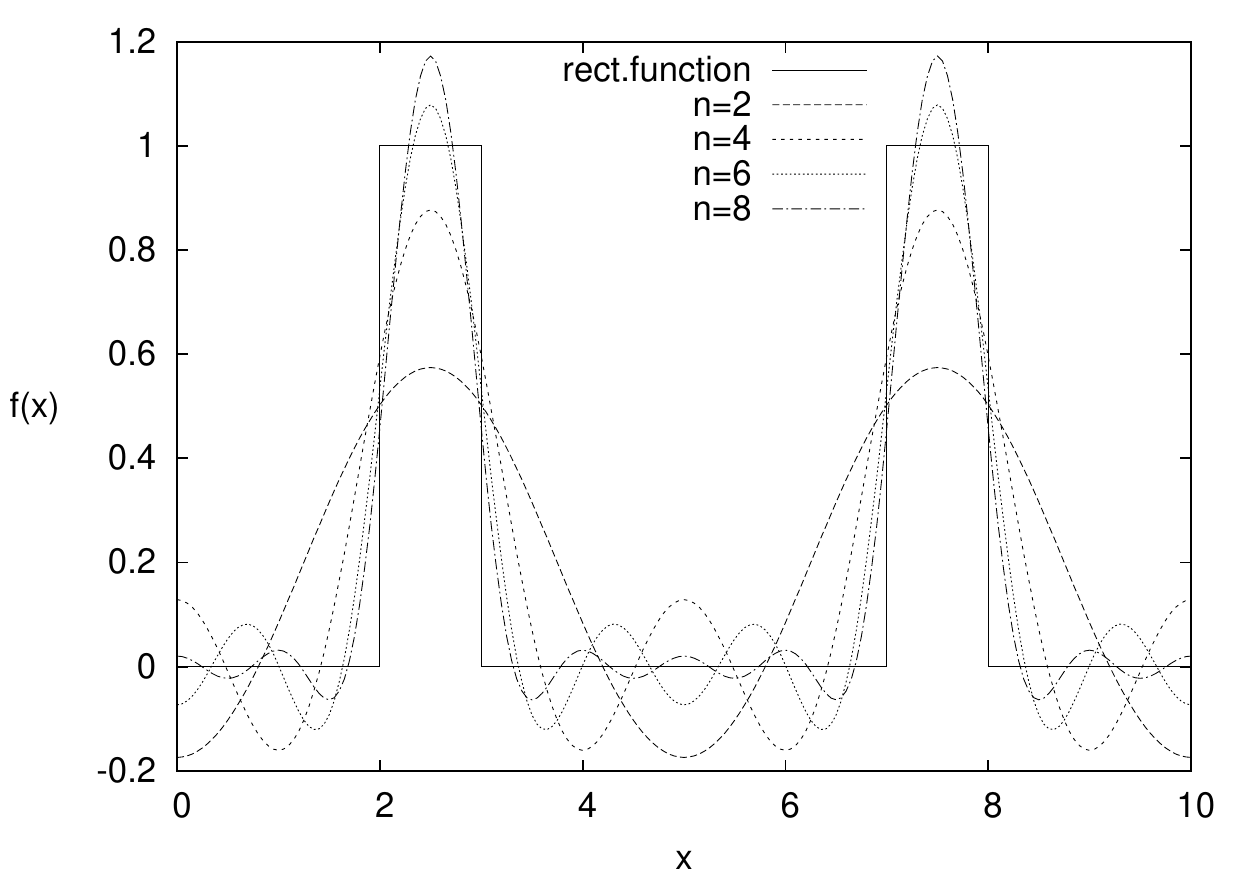}
	\caption{Fourier distribution of $\mathcal{P}_t=\{3, 8\}$ in document $D$, using different Fourier orders.} 
	\label{word_positions} 
\end{figure}

Figure \ref{word_positions} shows the Fourier representation of the step function $f^{(t)}(x)$ for the positions $\mathcal{P}_t=\{3, 8\}$ of the term \textit{t} in document $D$, calculated for different Fourier orders $n=2,4,6,8$.

\section{Comparing Term Distributions}
The underlying concepts of the proposed approach are:
\begin{itemize}
\item The positions of content terms in a document influence its relevance evaluation in the retrieval process.
\item If two content term distributions are similar, then the corresponding terms are located in a similar document region, implying some semantic relationship between them \cite{Kaszkiel97,Attar77,Tao07}.
\item The algorithm to compare two term distributions has to be computationally simple such that it can be performed under realistic conditions.
\end{itemize}

We argue that finite order Fourier approximations provide a systematic way to characterize and analyze the positions of terms. Applying a Fourier approximation of order $n$ reduces the data necessary to describe the term distribution to $2n + 1$ real numbers.

\begin{figure}[ht]
\centering
\includegraphics[width=6cm,bb=0 0 260 193]{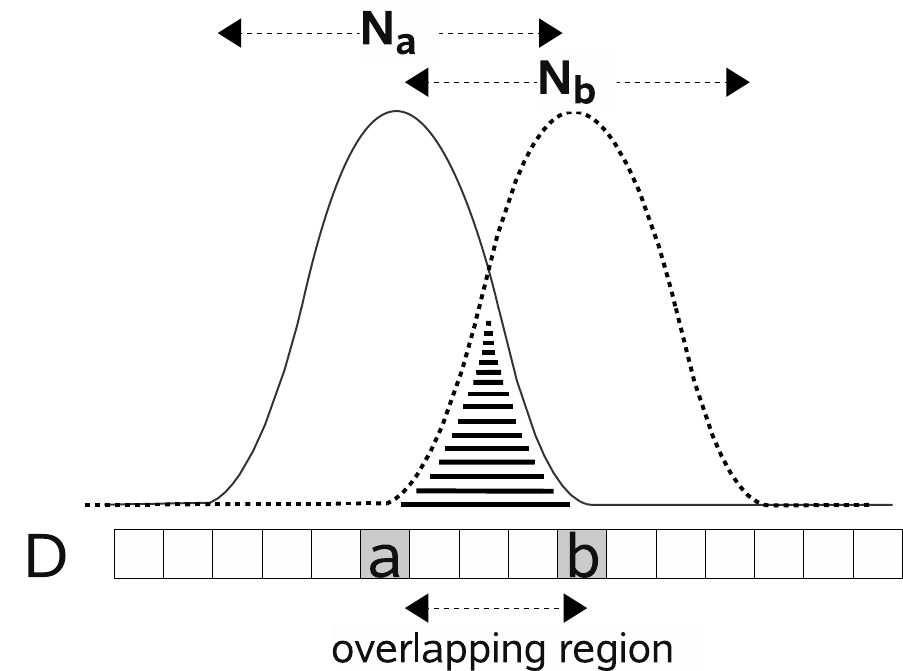}
  \caption{The broadening of the approximated term distributions, defining the term neighborhoods $\mathrm{N_a}$ and $\mathrm{N_b}$ and the corresponding overlapping region.}
\label{smear-out}
\end{figure}

In addition, the finite approximation allows to exploit the broadening effect on the original function (Figures \ref{word_positions}, \ref{smear-out}), defining a certain neighborhood around each term position. This broadening effect provides an instrument for estimating the similarity between terms within a document.

\subsection{Comparing the Term Distribution Functions}
In this section, the notion of {\em similarity\/} of two term distributions is defined. For a term distribution $f(x)$, the $n$-th order
Fourier approximation $f_n(x)$ is considered and its Fourier coefficients are used to form the $2 n + 1$ dimensional real vector $\vec{f}_n = (a_0, a_1, b_1, \ldots, a_n, b_n)$.

The similarity of two term distributions can be related to the overlap
integral
\begin{equation}
\label{equation_overlap}
\langle f_n, f_n' \rangle = \int\limits_0^L f_n(x) f_n'(x) \, \md x 
\end{equation}
The overlap integral measures in which regions of the integration range both functions are large (see Figure \ref{smear-out}). An important property of the Fourier expansion (\ref{fourier_expansion}) is that the overlap integral can be easily expressed by the spectral vectors $\vec{f}_n$ and $\vec{f}_n'$:
\begin{equation}
\langle f_n, f_n' \rangle = a_0 a_0' + \sum_{k = 1}^n (a_k a_k' + b_k b_k')
= \vec{f}_n \cdot \vec{f}_n^{\,\prime}
\end{equation}
i.e.\ the overlap integral is just the scalar product of the spectral
vectors \cite{Schaum74}.
Since the functions $f$ and $f'$ can represent terms from documents of different lengths, the overlap integral (\ref{equation_overlap}) is not used directly to define the similarity of term distributions, but instead the overlap of the {\em normalized term distributions\/} $f_n / \sqrt{\langle f_n, f_n \rangle}$ is used. It is simply the cosine of the angle between the spectral vectors:
\begin{equation}
\simil(f_n, f_n') = \cos \theta 
= \frac{\vec{f}_n \cdot \vec{f}_n^{\,\prime}}{|\vec{f}_n| |\vec{f}_n^{\,\prime}|}
\end{equation}
Here, the length of the spectral vector is given by
\[
|\vec{f}_n| = \sqrt{a_0^2 + \sum\nolimits_1^n (a_k^2 + b_k^2)}
= \sqrt{\langle f_n, f_n \rangle}
\]
\subsection{Relevance Ranking Optimization}
We state the document ranking problem as an optimization problem that is based on the query term distribution function $f_{q,d}$ and a \textit{user defined objective function} $f_\mathrm{o}$ representing the optimal query term distribution in the document body:
\begin{equation}
\label{objective_ranking_function}
Maximize\left\{ \simil(f_{q,d},f_\mathrm{o}) \right\} \hspace{1cm} \forall f_{q,d} \in A
\end{equation}
where $A$ represents the set of query term distributions in an initial document ranking, $f_{q,d}$ is the query term distribution function for query $q$ in document $d$, and $f_\mathrm{o}$ is a user defined \textit{objective function}, representing the \textit{optimal} query term distributions for the documents in the ranking.

For queries consisting of multiple terms, the distribution function is the sum of the single query term distributions.

Applying expression (\ref{objective_ranking_function}), a new sorted set of documents with a \textit{maximum} similarity between each document distribution $f_{q,d}$ and the objective function $f_\mathrm{o}$ is obtained. In other words, we get a new ranking in which the searched terms are distributed similarly to the optimal query term distribution described by $f_\mathrm{o}$.

\begin{figure}[ht]
	\centering
	\includegraphics[width=\linewidth,bb=0 0 751 186]{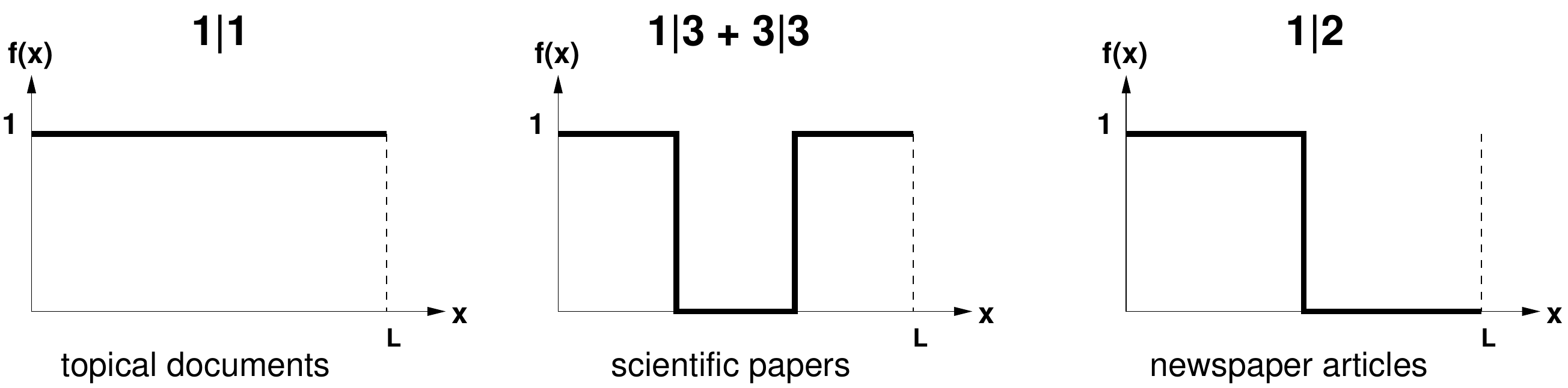}
	\caption{Examples of objective functions}
	\label{objective_functions}
\end{figure}

Figure \ref{objective_functions} illustrates several basic objective functions to identify documents where query terms are distributed in particular document regions. 
The following nomenclature is used to define an objective function:
\begin{definition}
\label{objective_function_definition}
The expression ``$f_\mathrm{o}:X|Y$'' represents an objective function to evaluate the relevance of documents with respect to the position of specific terms. Each document is divided into $Y$ equally sized sections of length $\frac{L}{Y}$. The terms situated in the $X^{th}$ section increase the document's relevance in the ranking.
\end{definition}
For example, the objective function $f_\mathrm{o}:1|1$ can be used to search for documents in which content terms (keywords) are distributed within the whole document body. It allows to identify so-called \textit{topical} documents \cite{Katz96}, where multiple keyword instances (\textit{topical terms}) represent the intensity with which a concept is treated within the document.

More sophisticated objective functions, such as $f_\mathrm{o}:1|2$ and $f_\mathrm{o}:1|3$ + $3|3$, can be used if the user is interested in documents where the contents of the first, or the first and the last section is more relevant. An example is the search for scientific papers where the abstract, the introduction (first sections) and the conclusion (last section) typically contain the most condensed document information. Another example might be a newspaper article, where readers expect to find the most relevant information at the top of the document.

Comparing the term distribution of our sample document $D$ (Figure \ref{pulse_distribution}) to Figure \ref{objective_functions}, it can be observed that $D$ will be only considered as relevant if the applied objective function resembles the pattern $f_\mathrm{o}:1|3$ + $3|3$.

\subsubsection{Algorithmic Complexity and Index Representation}
Each term distribution function (i.e.\ their Fourier coefficients) can be obtained using an algorithm
with a complexity of $O(\eta)$, where $\eta = termFrequency * fourierOrder$, and it will typically be executed in indexing time.

The most efficient index structure for text query evaluation is the \textit{inverted file}: a collection of lists (one per term) recording the identifiers of the documents containing that term \cite{Baeza-Yates99}. An inverted file index consists of two main components: a \textit{vocabulary} and a set of \textit{inverted lists}. The \textit{inverted lists} are represented as sequences of ${<}d,\nu_{d,t}{>}$ pairs, where $\nu_{d,t}$ is the frequency of term $t$ in document $d$. This is the standard document-level index in which term positions within documents are not recorded. 
In the proposed approach, this index is augmented with Fourier coefficients:
\begin{equation}
{<}d, a_0^{(t)}, a_1^{(t)},b_1^{(t)},\ldots,a_n^{(t)},b_n^{(t)}{>}
\label{formula_compresed_index}
\end{equation}
where $n$ is a predefined Fourier order and $a_k^{(t)}, b_k^{(t)}$ are the Fourier coefficients representing the positions of term $t$ in document $d$. Note that from (\ref{fourier_coef_0}), the component $a_0^{(t)}$ corresponds to the term frequency $\nu_{d,t}$.

The Fourier coefficients are computed by the indexing process. It should be emphasized that at query time these coefficients will be used to evaluate the similarity score between terms, by applying a simple scalar product calculation. We call this method Fourier Vector Scoring
(FVS).

\subsubsection{An Example}
Let us consider three arbitrary documents from the TREC-8 document collection containing the term ``\textit{brasil}''. The corresponding term distribution functions will now be compared with different objective functions, simulating two particular ranking criteria.

In Table \ref{simrank}, the similarity for each document using the Fourier order $n=3$ is shown. The applied objective function directly influences the ranking configuration, obtaining the documents FT944-15312 and FT931-11717 with the higher similarity (relevance) values for $f_\mathrm{o}:1|2$ and $f_\mathrm{o}:1|1$, respectively.

\begin{figure}[ht]
	\centering
	\includegraphics[width=1.0\linewidth,bb=0 0 350 252]{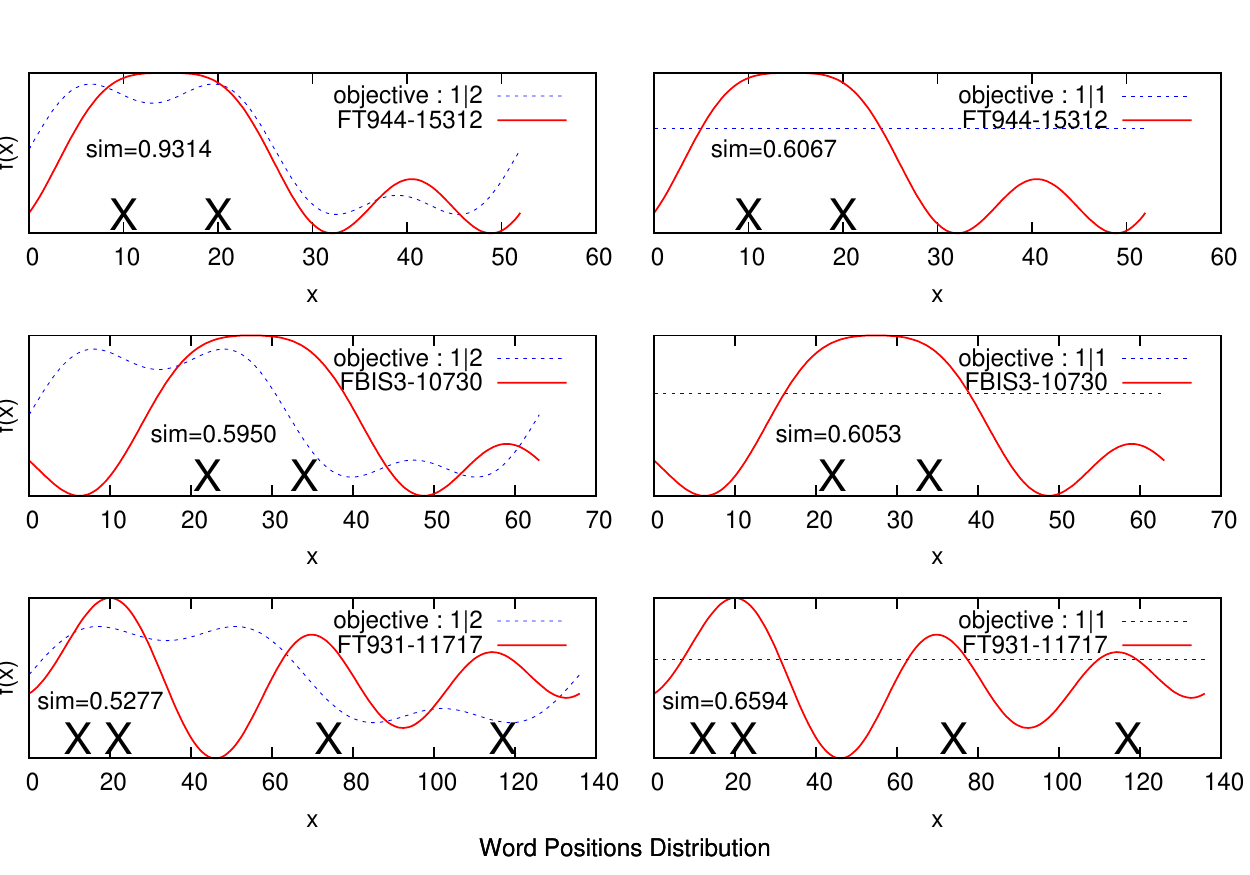}
\caption{The distribution of the term ``\textit{brasil}'' in three TREC documents, applying the objective functions: $f_\mathrm{o}:1|2$ (left) and $f_\mathrm{o}:1|1$ (right).}
\label{rank_comparison}
\end{figure}

Figure \ref{rank_comparison} indicates how documents whose term distribution approximates the applied objective function obtain a higher similarity value. For example, document FT944-15312 with $f_\mathrm{o}:1|2$ obtains a similarity value of $0.9314$, while the same document evaluated with $f_\mathrm{o}:1|1$ has a similarity value of $0.6067$, lowering its relevance in the ranking.
\begin{table}
\centering
\caption{Similarity and ranking for the query ``\textit{brasil}'' and three arbitrary TREC documents using the objective functions: $f_\mathrm{o}:1|2$ and $f_\mathrm{o}:1|1$.}
\small
\begin{tabular}{|c|c|c|c|c|}
\hline
\textbf{document} & \multicolumn{2}{|c|}{$f_\mathrm{o}:1|2$} & \multicolumn{2}{|c|}{\textbf{$f_\mathrm{o}:1|1$}} \\
\cline{2-5}
  & \textbf{sim} & \textbf{rank} & \textbf{sim} & \textbf{rank}\\
\hline
FT944-15312 & 0.9314 & 1 & 0.6067 & 2 \\
\hline
FBIS3-10730 & 0.5950 & 2 & 0.6053 & 3 \\
\hline
FT931-11717 & 0.5277 & 3 & 0.6594 & 1 \\
\hline
\end{tabular}
\label{simrank} 
\end{table}

\subsection{Query Expansion}
Query expansion (or term expansion) is a process of supplementing the original query ($q$) with additional terms, with the aim of improving retrieval performance \cite{Efthimiadis1996,Billerbeck03}.
The use of query expansion strategies such as automatic local analysis typically has positive effects on the retrieval performance.
Based on this observation, a new approach for query expansion is proposed, considering the \textit{top-r} documents $D=\left\lbrace d_1, d_2,\dots,d_r\right\rbrace$ of an initial ranking process. 

The function $f_{q,d}$ represents the distribution of the query term $q$ for each document $d$ $\in$ $D$. The set of terms $T_q$ whose elements $t$ maximize the expression $\simil(f_{q,d},f_{t,d})$ is computed. Using this expression, the terms for all documents in $D$ that have a similar distribution as the query, i.e.\ terms positioned near the query in the top ranked documents, are obtained.

\begin{figure}[ht]
\centering
\includegraphics[width=1.0\linewidth,bb=0 0 360 252]
{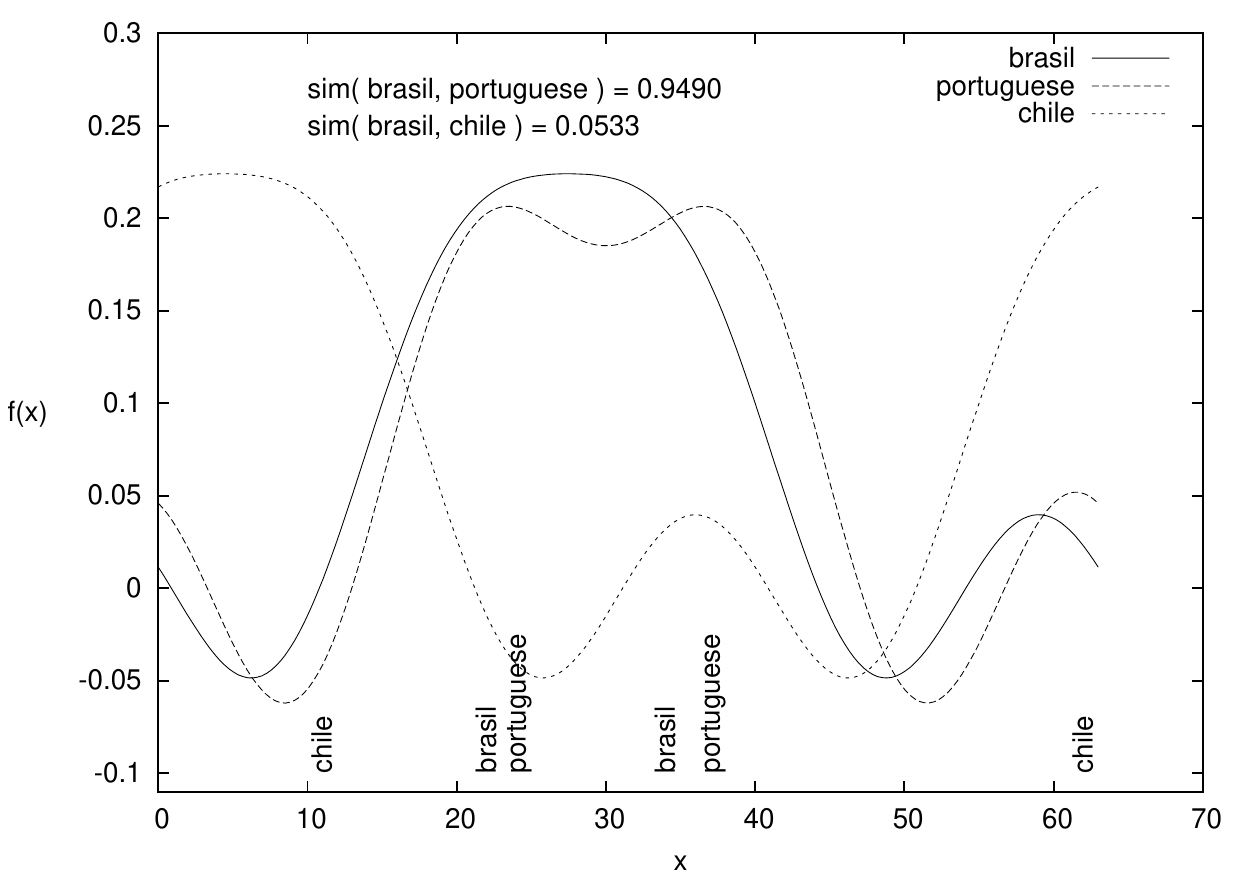}
\caption{Term neighborhood analysis for a typical TREC-8 document.}
\label{query-expansion-example}
\end{figure}

Taking a look at the term positions of a typical TREC-8 document (see Figure \ref{trecdoc-FBIS3-10730}),
it can be observed how the similarity criterion reflects the location properties of \textit{distant} and \textit{neighboring} terms (see Figure \ref{query-expansion-example}). The term ``\textit{brasil}'' and its neighbor term ``\textit{portuguese}'' have a high similarity value of $0.9490$, while its similarity value with respect to the more distant term ``\textit{chile}'' decreases to $0.0533$, which is about 20 times smaller. Thus, the proposed method is quite sensitive with respect to the location properties of terms.
\begin{figure}[ht]
\begin{scriptsize}
\begin{verbatim}
<DOC>
<DOCNO> FBIS3-10730 </DOCNO>
<HT> "drlat048_n_94005" </HT>
<HEADER>
<AU> FBIS-LAT-94-048 </AU>
Document Type:Daily Report 
<DATE1> 11 Mar 1994 </DATE1>
</HEADER>
<F P=100> Chile </F>
<H3><TI> Brazil's Franco Completes Schedule 
Despite Flu </TI></H3>
<F P=102> PY1103004294 Brasilia Voz 
do Brasil Network in Portuguese 
2200 GMT 10 Mar 94 </F>
<F P=103> PY1103004294 </F>
<F P=104> Brasilia Voz do Brasil Network </F>
<TEXT>
Language: <F P=105> Portuguese </F>
Article Type:BFN 
[Text] Although he has the flu and a fever of 38 degrees 
centigrade, President Itamar Franco is carrying out all 
commitments included on the agenda of his visit to Chile. 
</TEXT>
</DOC>
\end{verbatim}
\end{scriptsize}
\caption{A typical TREC-8 document.\label{trecdoc-FBIS3-10730}}
\end{figure} 

The expanded query is the set
\begin{equation}
 T^k_q=\{ \tau_1,\tau_2,\dots,\tau_k \}
\end{equation}
consisting of the $k$ best related query terms in $D$, obtained by ranking the terms according to the expression
\begin{equation}
\simil(f_{q,d},f_{\tau_i,d}),  \,  \forall \, d \in D, \, \tau_i \not= q 
\label{expanded_query_expression}
\end{equation}

The maximization process requires a simple comparison using the scalar product and norm of the corresponding Fourier coefficients, i.e.\ the algorithm to calculate the expanded query terms has a computational complexity of $O(\eta)$, where $\eta = |D|\,m + m\log m$, and $m$ is the number of terms in each document in $D$.

\section{Experimental Results}
The TREC-8 document collection has been used to measure the performance of the proposed approach. The goal of this evaluation is to determine how well the algorithm is able to identify documents based on a predetermined objective function, and to compare the proposed query expansion approach with some of the state-of-the-art models.

The evaluation framework consists of the following components: (a) the Ad hoc Test Collection containing 556,077 documents (2.09 Gigabytes) corresponding to the Tipster disks (3 and 4), (b) the Topics and Relevance Judgments (qrels), (c) our approach consisting of 4 Java modules for indexing, search, graphical evaluation and configuration tasks, and (d) the results analysis where the effectiveness of our approach will be estimated.

\subsection{Objective Function Runs}
In this experiment, it will be analyzed how the query terms (TREC topics) are distributed in the \textit{top-10} ranked documents for three different ranking schemes: (a) tfidf (baseline) and two objective functions: (b) $f_\mathrm{o}:1|3$ and (c) $f_\mathrm{o}:3|3$.

\begin{figure*}[ht]
	\centering
	\includegraphics[width=1.0\linewidth,bb=-40 0 440 332]{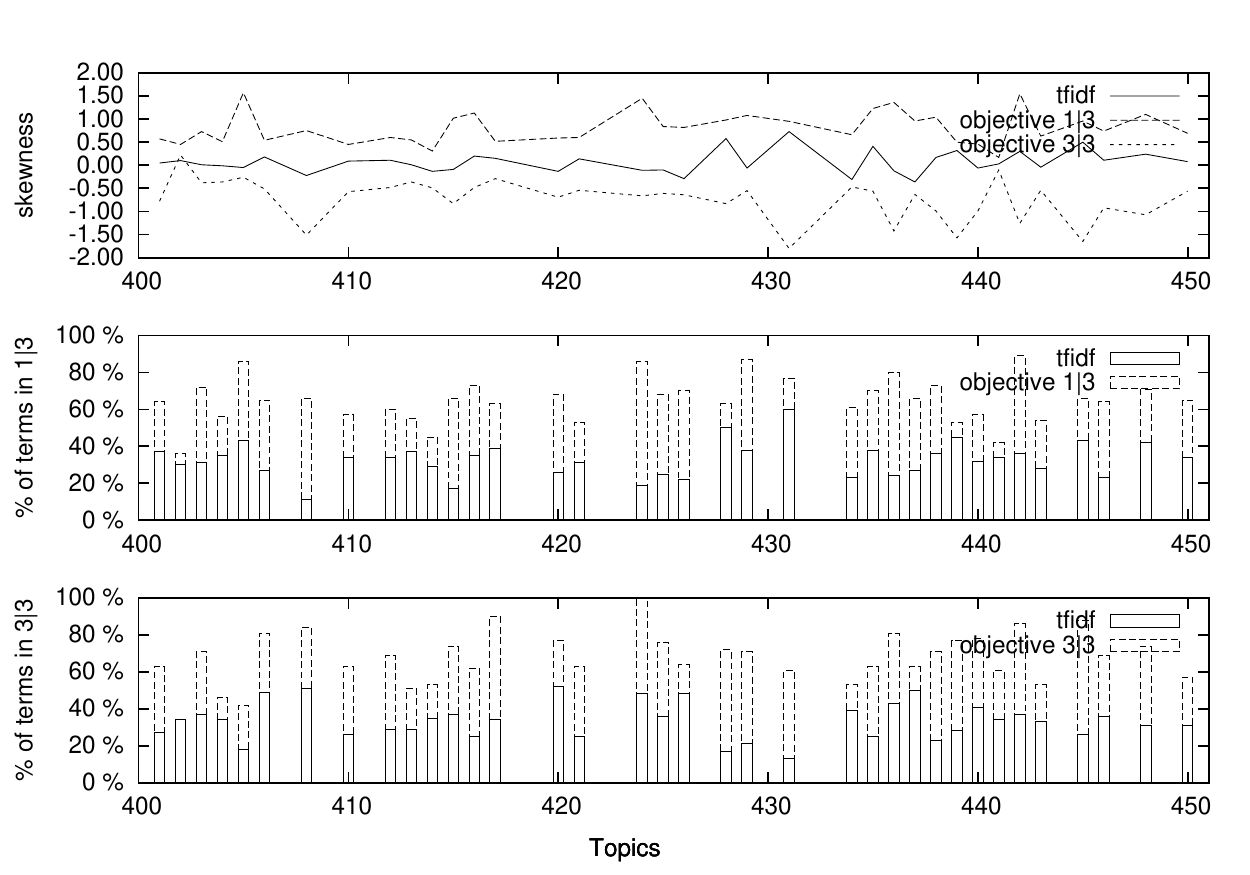}
	\caption{Analyzing the query term distribution skewness for two different ranking schemes: $f_\mathrm{o}:1|3$ and $f_\mathrm{o}:3|3$.}
	\label{ranking_skewness}
\end{figure*}

To measure how the developed ranking algorithm follows the proposed objective functions, the skewness \cite{Moore03} of the term position distributions is calculated and their asymmetry is compared with the \textit{tfidf} scheme. To obtain relevant statistical results, only topics that return more than 10 document hits were considered.

As depicted in the first graph of Figure \ref{ranking_skewness} (skewness), the query term distribution in the \textit{tfidf} ranking has a skewness of around zero, i.e. the terms are evenly distributed.
Applying both objective functions, it can be observed how in the optimized ranking the query terms approximate the corresponding objective function: the ranking based on $f_\mathrm{o}:1|3$ shows a positive skewness, demonstrating that terms are mainly situated in the header of the ranked documents. On the other hand, applying $f_\mathrm{o}:3|3$ generates a document ranking where query terms are predominantly distributed at the document's bottom (negative skewness).

The last two graphs of Figure \ref{ranking_skewness} show the rate of query terms fitting the proposed objective functions. For example, the $f_\mathrm{o}:1|3$ function applied to \textit{topic} 420 produces a document ranking where 68\% of the query terms are situated inside the objective function region, while the \textit{tfidf} ranking returns only a fitting rate of 26\%.
Analyzing all experimental results, it can be observed that by applying the proposed approach to TREC-8, about 67\% of the query terms (from the top ranked documents) are positioned inside the defined objective function region.
Therefore, it is evident that the ranking process can be flexibly optimized, providing new possibilities to express the  information need of the user.

\subsection{Query Expansion Runs}
Our query expansion experiments are based exclusively on the search results.  No external knowledge structure was used to leverage the re-ranking procedure.

In the group of runs described in the following, the proposed query expansion model based on the query terms distribution ($f_q$) is evaluated. 

Using the top-\textit{n} ranked documents, the query distribution function $f_q$ for each ranked document is obtained, and the terms having a  similar distribution as $f_q$ are calculated. Based on equation (\ref{expanded_query_expression}), the first $k$ candidate terms $T^k_q=\left\lbrace \tau_1,\tau_2,\dots,\tau_k \right\rbrace $ for query reformulation are obtained and the new ranking using our test collection is evaluated.

The query reformulation and ranking procedure consists of the following steps:
\begin{enumerate}
\item Calculate the expanded query terms $T^k_q$ based on the top-\textit{n} documents from the \textit{tfidf} ranking.
\item Using $T^k_q$, calculate the expanded query
\begin{equation}
q_\mathrm{e}  = \left\lbrace w_0 q, \; w_1 \tau_1, \; w_2 \tau_2, \; \dots, \;  w_k \tau_k \right\rbrace \\
\label{formula_expanded_query}
\end{equation} 
where $w_i$ is a weighting factor corresponding to the similarity between the original query $q$ and the term $\tau_i$.
\item Perform the \textit{tfidf}-search with $q_\mathrm{e}$.
\end{enumerate}

Using the top-10 ranked documents and the first 40 terms having the highest query similarity, the proposed Fourier Vector Scoring (FVS) query expansion method for $w_i=1$ is compared with eight state-of-the-art query-expansion methods: Rocchio for $\beta=0.2$, 0.4, 0.6, 0.8, 1.0 (Ro.2, Ro.4, Ro.6, Ro.8, Ro1) \cite{Rocchio71}, Bose-Einstein 1 (Bo1), Bose-Einstein 2 (Bo2) \cite{Amati02} and Kullback-Leibler (KL) \cite{Cover91}.  For the query expansion experiments, the Terrier \cite{Ounis06} software was used.

\begin{figure}[ht]
 \centering
 \includegraphics[width=1.0\linewidth,bb=0 0 350 262]{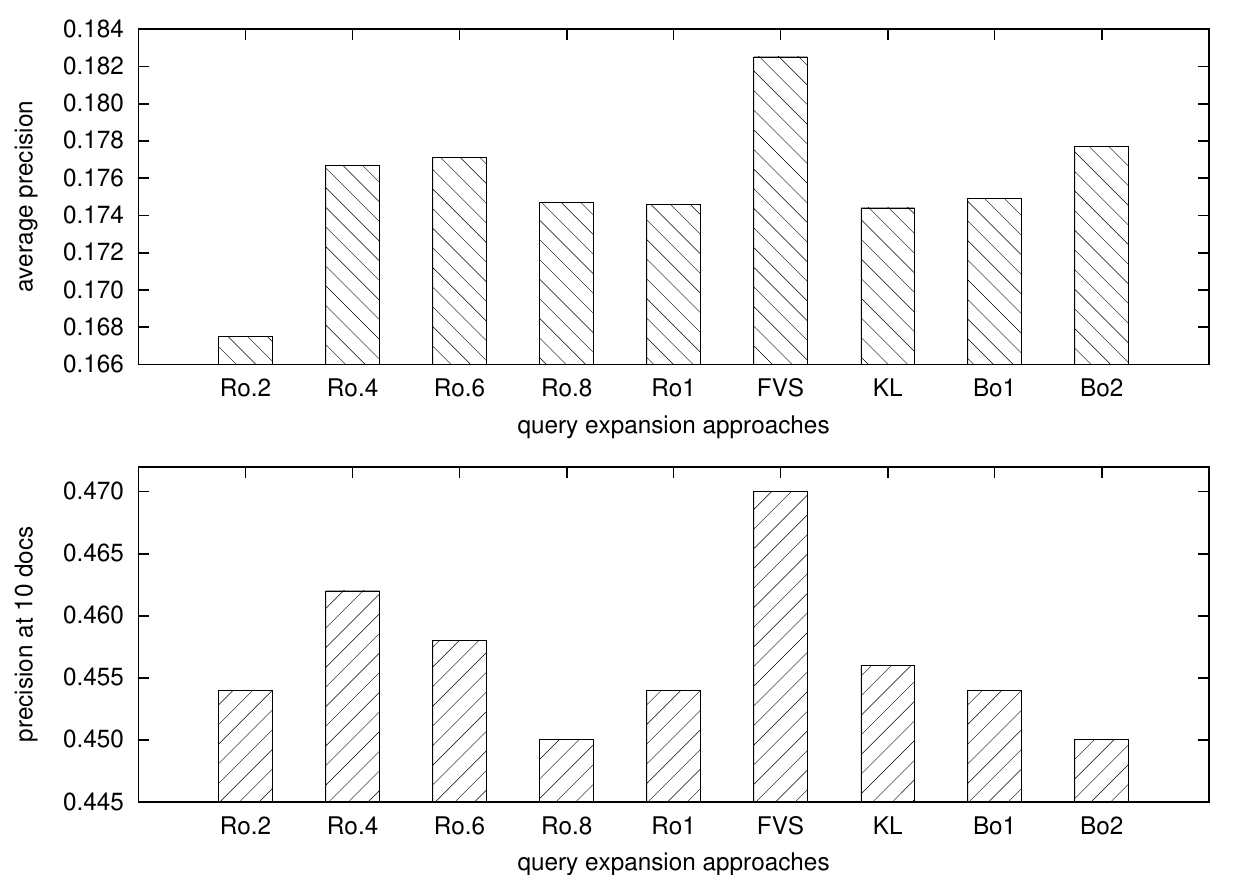}
 \caption{Ranking improvements using query expansion.}
 \label{expanded_query_graph}
\end{figure}

Considering the measures of \textit{relevance precision} and \textit{precision at 10 documents}, it can be observed from Figure \ref{expanded_query_graph} that FVS outperforms all other query expansion methods.

Table 2 shows the most relevant expanded terms, listed in descending relevance order, for eight arbitrary topics from the test collection. The same term sets were also used in the query expansion runs.

\begin{table}
\caption{Examples of query expansion terms for some arbitrary TREC-8 Topics.}
\scriptsize
\begin{tabular}{|c|p{0.25\linewidth}|p{0.50\linewidth}|}\hline
\textbf{Topic} & \textbf{Title Query} & \textbf{Terms for Query Expansion}\\
\hline
403 & osteoporosis & bone, women, calcium, health, risk, study, claim, research\\
\hline
406 & Parkinson's disease & brain, research, cells, london, drug, symptoms, alzheimer, fetal\\
\hline
408 & tropical storms & july, disaster, area, caribbean, hurricane, texas, georgia, temperatures\\
\hline
417 & creativity &  people, mental, illness, scientists, part, human, children, depression\\
\hline
421 & industrial waste disposal & management, facilities, hazardous, radioactive, solid, company, state, site\\
\hline
427 & UV damage, eyes & radiation, rays, sunglasses, protect, adhesive, patch, exposure, children\\
\hline
429 &  Legionnaires' disease & nosocomial, hyph, infection, control, patients, prevention, pneumonia\\
\hline
431 & robotic technology & robot, manufacturing, industrial, system, company, human, industry\\
\hline
\end{tabular} 
\end{table}

\section{Conclusions}
In this paper, term distribution analysis using Fourier series expansion has been proposed as a novel methodology to improve document relevance evaluation in information retrieval applications.
The proposed approach is based on a Fourier series representation of the term positions in a document collection, by calculating the corresponding expansion coefficients.
By using query objective functions for predetermined document regions, the approach provides new ways to define or refine  queries.
Furthermore, a novel query expansion methodology has been presented to support the user in the query refinement process.

An evaluation of our proposal using the TREC-8 collection has demonstrated that 67\% of the query terms are positioned inside the user defined objective function region. A further analysis has shown that using the proposed approach to generate expanded query terms leads to a performance gain over state-of-the-art query expansion models such as Rocchio and \textit{Divergence from Randomness} models.

There are several issues for future work. For example, it would be interesting to study the possibility of generating optimized objective functions by training our approach with particular document categories such as medical, juridical, scientific papers, etc.
A further topic to be investigated is the compression of the index information, because the size required to save the spectral information of low frequency terms currently exceeds the size of the positional information of terms.
Finally, the presented approach is not limited to the Fourier expansion, but can be generalized to other kinds of series to represent the term distribution functions.

\bibliographystyle{abbrv}
\bibliography{jcdl2009}

\end{document}